# Development of EAP-based actuators for high-frequency adaptive optics system

A. Michel*[a], D. Audigier [a], C. Richard [a], J-F. Capsal [a]
[a]INSA Lyon, LGEF, UR682, Villeurbanne, France

## ABSTRACT

The present work aims to enhance the electrostrictive strain of the P(VDF-TrFE-CFE) terpolymer for use in adaptive optics, specifically in deformable mirror actuation. In the context of the FlexSiMirror project, these systems seek to operate under alternating electric fields of up to 50 V/µm and within the kilohertz (kHz) frequency range, thereby framing the ranges of characterization considered in this study. To achieve greater strains, the incorporation of a polymeric plasticizer up to 20 vol.% and its impact on the actuation strain performance was studied. Hence, the relevance of this approach lies both in the kHz characterization of the mechanical and dielectric properties of the materials and on the utilization of a polymeric plasticizer, instead of the commonly used phthalates suitable for the lower frequency ranges. In the kHz range, polymeric plasticizer addition markedly reduces the elastic modulus while limiting molecular migration, leading to more than a threefold increase of the figure of merit associated with strain ($FoM_{strain}$) and yielding 1.50 % strain output under 50 V/µm, 3.6 times greater than that of the unmodified terpolymer. Therefore, these findings show that modified P(VDF-TrFE-CFE) exhibits enhanced electromechanical performance in the kHz range. This advancement opens new possibilities for developing next-generation actuators intended for adaptive optics applications.

## 1. INTRODUCTION

Actuation performance under high frequency electric fields is being increasingly sought, especially in the context of adaptive optics where ultra-fast deformable mirror technologies are to be expected. In the FlexSiMirror project, funded by the National Research Agency (ANR), as part of the research program *France 2030*, a silicon wafer measuring 100 µm in thickness and 20 cm in width is supported by pillar-shaped actuators made of a field-activated electroactive polymer (EAP), as depicted in Figure 1. These actuators are being designed to enable localized mirror deformation within milliseconds corresponding to frequencies of alternative electric fields in the kilohertz (kHz) range. The longitudinal deformation of a single pillar is anticipated to be on the order of a few micrometers (2-10 µm), with crosstalk deformation estimated to be 30% of this value when exposed to electric fields up to 50 V/µm.

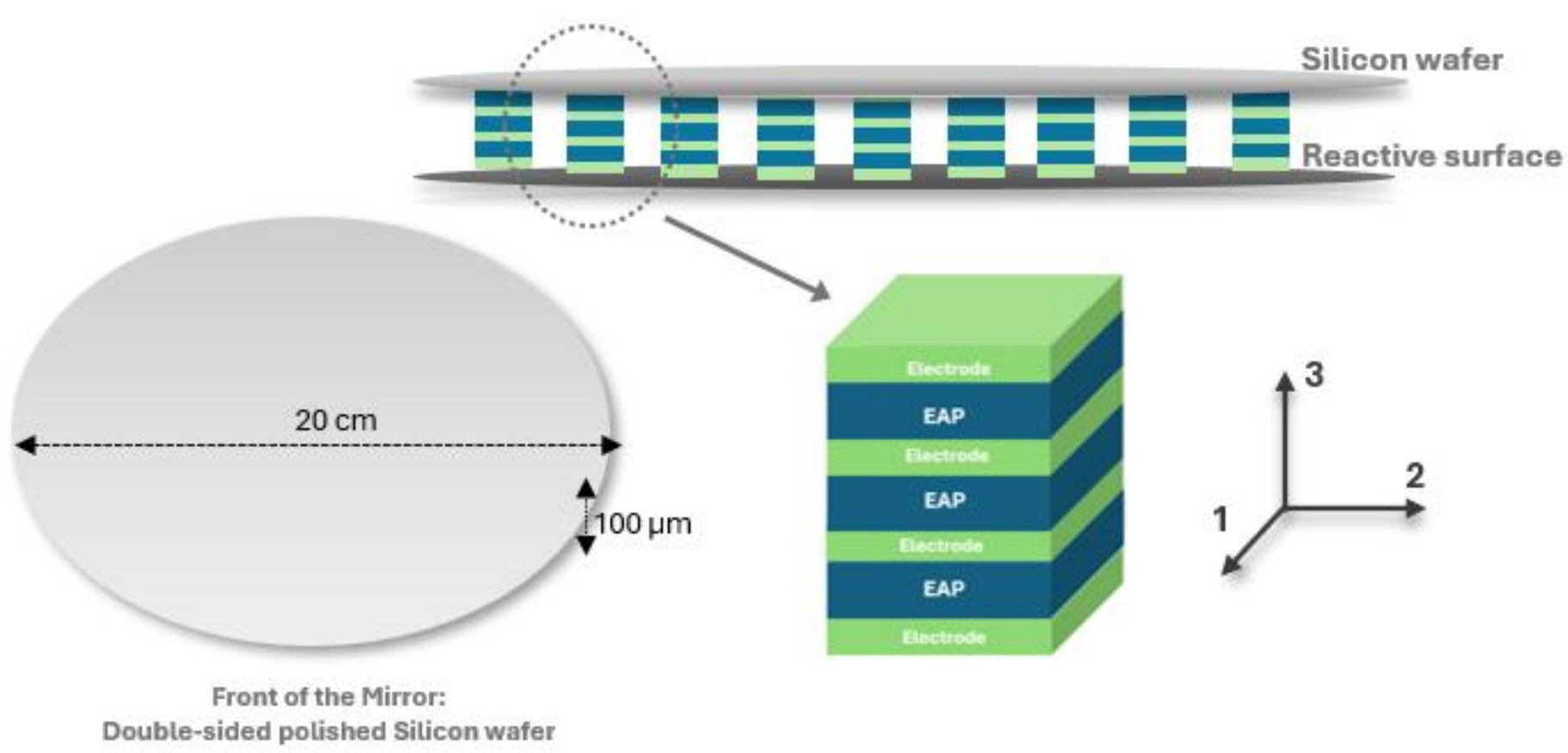

*Figure 1: FlexSiMirror actuator system; silicon wafer specifications on the left side, actuator in "pillar" shape of the right side. Actuation electric field is applied within direction (3).*

*audrey.michel@insa-lyon.fr

Although piezoelectric materials like the copolymer P(VDF-TrFE) are often used as actuators, their requirement to be poled at electric fields near 100 V/µm to reach their ferroelectric state [1] limit their suitability for multi-layer actuator architectures. Poling a large number of stacked layers is difficult to achieve without risking dielectric breakdown in some of them. This consideration motivates the present study to focus instead on materials that do not require poling, specifically electrostrictive terpolymers such as P(VDF-TrFE-CFE/CTFE). The latter have attracted considerable attention because they offer competitive actuation performance, especially regarding their strain output at relatively low electric fields (≤50 V/µm) thereby substantially reducing the likelihood of dielectric breakdown, which is typically reported to occur above 140 V/µm [2]. When subjected to an electric field in the direction (3) along the thickness, the electrostrictive film shrinks longitudinally and expands laterally, yielding a negative strain in the direction (3) ($S_{33}$) and a positive one in the direction (1) ($S_{31}$) respectively. Below fields of 100–150 V/µm, these strains mostly result from Maxwell's electrostatic pressure [3] acting between the upper and lower surfaces of the electrostrictive film and are classically described as follow:

$$S_{33}(E) = \frac{\varepsilon(E)}{2Y}(1+2\mathrm{v})\,E^2, \quad (1) \qquad S_{31}(E) = \frac{S_{33}(E)}{\mathrm{v}}; \quad (2)$$

with $\varepsilon(E) = \varepsilon_0\varepsilon'(E)$ the dielectric permittivity, Y' the elastic modulus and v the Poisson's ratio of the terpolymer.
Nellie Della Schiava et al., established that for such electrostrictive material, actuation performances can be assessed through figures of merit relative to strain, blocking force ($T_b$) and mechanical energy density ($W_m$), strongly dependent on the material's permittivity and elastic modulus, expressed as follows:

$$FoM_{strain} = \frac{\varepsilon_0\varepsilon'}{Y'} \quad (3) \qquad FoM_{T_b} = \varepsilon_0\varepsilon' \quad (4) \qquad FoM_{W_m} = \frac{(\varepsilon_0\varepsilon')^2}{Y'} \quad (5)$$

Over the last decade, the incorporation of low molecular weight plasticizers also known as phthalates, such as DINP or DEHP, has demonstrated an improvement of the electromechanical performance of terpolymers for frequencies up to the Hertz range [4]. CFE-based terpolymers combined with DEHP in particular showed a significant improvement of $FoM_{Strain}$ , greater than CTFE-based terpolymers [5], orienting this work on PVDF-TrFE-CFE terpolymer. However, the use of phthalates is increasingly restricted by environmental regulations due to concerns about their long-term toxicity [6]. In this context, polymeric plasticizers are proving to be a particularly attractive alternative. Unlike DINP or DEHP, polymeric molecules have much better chemical stability and are much less prone to migration and leaching out of the matrix, which prevents premature degradation of the actuator's properties [7]. These characteristics are all the more crucial in adaptive optics requiring operating frequencies up to the kHz range. Indeed, while conventional plasticizers excel at very low frequencies (100 mHz - 30 Hz) by promoting interfacial polarization also known as Maxwell-Wagner-Sillars (MWS) effect, they also generate significant ionic conduction losses. The use of a polymeric plasticizer stabilizes chain dynamics and reduces these losses, offering a better compromise for applications requiring wider bandwidth and long-term electromechanical stability [4] [7].

This work will first concentrate on the characterization of the dielectric and mechanic properties of the neat and modified terpolymers in the kHz range and under electric fields up to 50 V/µm. The objective is to measure the physical quantities associated with $FoM_{strain}$ as well as the electric field threshold for the actuation of such materials. Subsequently, an analysis of the resulting FoMs and the corresponding reachable strain outputs will be performed to assess the degree of actuation enhancement achieved through the incorporation of a polymeric plasticizer.

## 2. MATERIAL PREPARATION AND CHARACTERIZATIONS

### 2.1 Materials

Poly(vinylidene-trifluoroethylene-chlorofluoroethylene) [P(VDF-TrFE-CFE)] powder with a composition of 49.5/37.5/12.9 was provided by Piezotech (Arkema Group, France). The material was synthesized through suspension polymerization.
Methyl-ethyl-ketone (MEK; AnalaR NORMAPUR Reag. Ph. Eur., ACS, 100.0%) was sourced from VWR Chemicals, cyclopentanone (purity >99%) from Carl Roth, and Durapore 5 µm PVDF membrane filters from Merck (Millipore).
Low viscosity polymeric plasticizer Palamoll 652 was provided by BASF.

### 2.2 Film elaboration and processing

Terpolymer powder was dissolved in a 60:40 mixture of MEK and cyclopentanone, with a polymer-to-solvent mass fraction of 16%. After formulation, the solutions were filtered through a 5 µm PVDF membrane. Different contents of Palamoll 652 were added, leading to six different solution formulations: pure terpolymer, terpolymer with 5, 10, 15, 17, and 20 vol.% of Palamoll.

The solutions were then cast onto glass substrates using the doctor blade technique (elcometer 3700) and left at room temperature for 24 hours to allow the solvents evaporation. The resulting six films were then placed in a convection oven at 60°C for 2 hours to ensure complete removal of any remaining solvent, followed by a 90°C annealing for an additional 2 hours.

### 2.3 Infrared (IR) Spectroscopy

Infrared (IR) spectroscopy using an ATR module with a diamond crystal on a ThermoFischer FTIR Nicolet iS50 was conducted on plasticized films to determine the accurate Palamoll volume content right after the film elaboration and to estimate the plasticizer retention over time. For each sample, an average of 32 spectra was recorded in the range of 525 to 4000 $cm^{-1}$ with KBr optics at room temperature.

The volume contents of the films were determined by comparing the areas of a strong peak characteristic of Palamoll located at 1730 $cm^{-1}$. This analysis remained comparative, with the assumption that the area of the peak observed in the film containing 15 vol.% of Palamoll accurately reflects an actual content of 15 vol.%. The margin of error for the volume content of the plasticized films was determined by calculating the mean standard error from six repeated measurements conducted on the same sample.

### 2.4 Dielectric characterization at 1 kHz

Dielectric properties of neat and modified terpolymers were characterized first over a wide frequency range and under low electric fields (0.01 V/µm) using broadband dielectric spectroscopy (BDS) and secondly at 1 kHz under 50 V/µm through unipolar polarization cycles coupled with Debye-Langevin model.

BDS was performed using Novocontrol impedance analyzer. Samples of 100 µm thick were electrode with gold using a Cressington 208 HR sputter coater. Each side was sputtered at 40 mA for 2*40 seconds, creating 8 mm diameter disc electrodes.

From BDS, it is possible to retrieve the real and imaginary parts of the permittivity, expressed as $\varepsilon^* = \varepsilon' - i\varepsilon''$ and the loss tangent $\tan(\delta)_d = \frac{\varepsilon''}{\varepsilon'}$ [8]. $\varepsilon'$ is the real relative permittivity, $\varepsilon''$ the imaginary part accounts for the dielectric loss factor and $\tan(\delta)_d$ quantifies the phase shift between the application of the electrical field and the polarization response of the samples.

The dielectric properties were collected at room temperature over a frequency range of $10^{-1}$ to $10^{6}$ Hz with eight measurements per decade at 1 $V_{rms}$.

Unipolar polarization cycles were conducted on the 8 mm diameter film electrodes plunged into an oil bath to prevent air discharge from impacting the current detection. An AC voltage corresponding to an electric field of 50 V/µm was applied to the electrodes through a Keysight 33500B signal generator, amplified by the Trek Model 20/20C high-voltage power amplifier, providing precise control of output voltages in the range of 0 to 20 kV DC with an output current range up to ±20 mA DC.

The resulting current delivered to the electrodes is amplified by the current amplifier SR570 from Stanford Research Systems, exhibiting a maximum input current of $\pm$ 5 mA and a sensitivity from 1 pA/V to 1 mA/V in a 1-2-5 sequence. The accuracy of the voltage output $V_{out}$ delivered by the amplifier is estimated to be equal to $V_{out} \pm (0.5\% \, V_{out} + 10\, mV)$.

The current response of the sample is transcribed on the oscilloscope Keysight EDUX 1002G, from where the data are exported and further analyzed.

The overall current density $J$ in a dielectric is defined as the sum of conduction losses $J_l$ and polarization current densities. In the case of neat and modified P(VDF-TrFE-CFE) at 1 kHz and under 50 V/µm, the polarization contribution results directly from the capacitive behavior of the material, defined by the capacitive current density $J_c$. Consequently, $J$ can be expressed as follows:

$$J = J_l + J_c \quad (6)$$

From $J_c$ it is possible to retrieve the real permittivity $\varepsilon'(E)$ of the materials under 50 V/µm as it originates from the electric displacement D, function of ε and $\frac{dE}{dt}$ and expressed as follows:

$$J_C = \frac{dD}{dt} = \varepsilon_0 \varepsilon' \frac{dE}{dt} \quad (7)$$

However, for semi-crystalline polymers such as P(VDF-TrFE-CFE), $\varepsilon'$ along with the saturation electric field $E_s$ arise from at least two contributions: the crystalline and the amorphous phases. Both the saturation field and the relative permittivity of the amorphous phase $E_s^a$ and $\varepsilon_a'$ are well established, reported to equal respectively 400 V/µm and 12 [9], [10]. These values are not restrictive when compared to the crystalline phase, which saturates at lower fields $E_s^c$, closer to the levels of the actuation operating range, causing a drop in $\varepsilon_c'$. To consider the dielectric disparities of such materials when subjected to 50 V/µm, the Debye-Langevin model is employed. Both $\varepsilon_c'(E)$ and $E_s^c$ can be obtained from a fit of the experimental curve $J_c(E)$ with the theoretical capacitive current computed through the Debye-Langevin model $J_{c_theo}$ [11] . $J_{c_theo}$ is a function of ε and $\frac{dE}{dt}$ but with $\varepsilon'(E)$ expressed as in Equation (8):

$$\varepsilon'(E) = 3(\varepsilon_c'(0) - 1)\left[\left(\frac{E_s^c}{E}\right)^2 - \left(\sinh\frac{E}{E_s^c}\right)^{-2}\right] + 3(\varepsilon_a'^{(0)} - 1)\left[\left(\frac{E_s^a}{E}\right)^2 - \left(\sinh\frac{E}{E_s^a}\right)^{-2}\right] \quad (8)$$

Dielectric losses $tan(\delta)$ at 50 V/µm and 1 kHz have been estimated from the phase shift Φ between the applied voltage and the current response of the samples undergoing polarisation cycles. Specifically, $tan(\delta)$ can be expressed as:

$$tan(\delta) = \frac{I_L}{I_c} = \frac{I * \cos(\Phi)}{I * \sin(\Phi)} \quad (9)$$

The main source of error comes from the graphic reading of the time lag with an accuracy of ± $10^{-5}$ s, leading to a ± 3.6 ° phase shift uncertainty and to less than 10% error on the $tan(\delta)$ values. In an ideal capacitor the current leads the voltage by exactly 90°, meaning the electric field stores and returns energy without dissipation [12].

The electrical breakdown strength $E_b$ was determined by applying DC voltage ramp at 500 V/s rate using a 50 kV high-voltage power supply: SR-50-N-500 model from Technix. The set-up is identical to that used for polarization tests, as previously described. Tests were performed on 16 samples of each film. $E_b$ can be estimated through the Weibull cumulative distribution function (CDF) defined as:

$$P(E) = 1 - \exp\left[-\left({}^{E}/_{\lambda}\right)^k\right] \quad (10)$$

With $P(E)$ the breakdown probability of the material for a specified E, $\lambda$ the scale parameter that reflects the breakdown strength at which 63.2% of the breakdown occurs, and $k$ the spread of the distribution. Higher $k$ values correspond to more uniform and reliable breakdown behavior [13]. One should bear in mind that the results are solely an estimation of P(E), giving an order of magnitude of the expected breakdown electric field $E_b$ for such materials. Still, to assess the quality of the Weibull CDF fit with the experimental data, the coefficient of determination $R^2$ is utilized and detailed in the results.

## 2.5 Mechanical characterization at 1 kHz

Dynamic Mechanical Analysis (DMA) using a Metravib 25 instrument was performed on the pure and plasticized films to characterize the Young Modulus $Y^*$ of the films, expressed as $Y^* = Y' + iY''$ with $\tan(\delta)_m = \frac{Y''}{Y'}$ for a viscoelastic material [14]. $Y'$ is known as the elastic modulus and $Y''$accounts for the loss modulus. Equally important is the damping factor $\tan(\delta)_m$ being the ratio between the dissipated and stored energies, indicating a strong viscous nature for values close to 1 against pure elastic behavior for values approaching 0.

The Time Temperature Superposition (TTS) principle coupled with the Wiliam-Landel-Ferry (WLF) model were used to predict the mechanical properties of the films outside of the experimental timescale [15]. TTS is a well-known principle for viscoelastic materials stating that the mechanical properties of polymers at high temperatures are equivalent to those at low frequencies, and vice versa. From $Y(f)$ measured by DMA at a reference temperature, here $T_0$=20°C, it is possible to shift the frequency isotherms of $Y'$ and $Y''$ measured at $T_i < T_0$ towards higher frequencies. These horizontal translations lead to the construction of master curves giving $Y'$ and $Y''$ over a wide frequency range ($10^{-1}$ to $10^5$ Hz). To ensure that the application of TTS principle is valid, a fitting of the master curve of $Y'$using the Arrhenius model if $T_0 < T_g$, with $T_g$ the glass transition, or the WLF model if $T_0 > T_g$ can be conducted. P(VDF-TrFE-CFE) exhibiting a $T_g$ around -20°C [16], WLF model, given in Equation (11) is considered.

$$\log(\alpha_T) = \frac{-C_1 \times (T - T_0)}{C_2 + (T - T_0)} \text{ , [17]} \quad (11)$$

with $\alpha_T$ is the shift factor, $C_1$ and $C_2$ are the WLF constants, $T$ is the measurement temperature and $T_0$ is the reference temperature. Universal values of $C_1$ and $C_2$ have been estimated at 17.44 and 51.6 K respectively and are said to be valid for reference temperatures chosen between $T_g$ and $T_g + 100\ °C$ [17]. The WLF model is applicable under the assumption that the material is thermorheologically simple, meaning that its relaxation mechanisms exhibit the same temperature dependence across all frequencies [15]. Because electrostrictive polymers exhibit a broad ferroelectric–paraelectric transition, peaking at the Curie temperature $T_c$, which reaches $-1.9$°C for the present P(VDF-TrFE-CFE) terpolymer [18], the validity of the WLF model across this transition may be questioned. However, this issue has been addressed by Y. Wang *et al.*, who demonstrated that the glass-transition–related α-relaxation remains the dominant relaxation mechanism from its maximum at $T_g$up to near room temperature for this terpolymer, despite a partial overlap with the broadened Curie transition [19].

The tensile specimens were 0.1 mm thick $(e)$, 10 mm wide $(l)$ and 30 mm high $(h)$ once clamped in the jaws, in accordance with the shape factor condition $h > \frac{2el}{e+l}$ and the maximal sizes cautioned by Metravib 25. Tensile tests were conducted using dynamic deformation rate at 0.1% and static deformation rate at 0.25%. Isothermal frequency sweeps were performed with seven points ranging from 0.1 to 50 Hz. Measurements covered temperatures from 20°C down to -20°C in 5°C increments, with a cooling rate of 2°C per minute.
The resulting curves of $Y'$ and $\tan(\delta)_m$ as a function of frequency ($10^{-1}$ to $10^5$ Hz) were plotted at $T_0$= 20°C for all samples. The margins of error of $Y'$ and $\tan(\delta)_m$ were estimated to be of ±5 and 10% respectively.

## 3. RESULTS AND DISCUSSION

### 3.1 Analysis of the Palamoll content over time

The IR Spectroscopy comparative analysis revealed the accurate volume contents of Palamoll within the plasticized films (terpolymer x Palamoll i vol.% hereafter referred to as xPALA i vol.%). To quantify the capacity of the terpolymer to retain the polymeric plasticizer over time, first measurements were taken at $t_0$ and second measurements were performed at $t_0$ + 120 days. The spectra centered at 1730 1/cm are given on Figure 2 and the results are summarized in Table 1. An initial observation drawn from Table 1 indicates that the first measured Palamoll contents are in close agreement with the target formulations, thereby validating the accuracy of the preparation procedure.

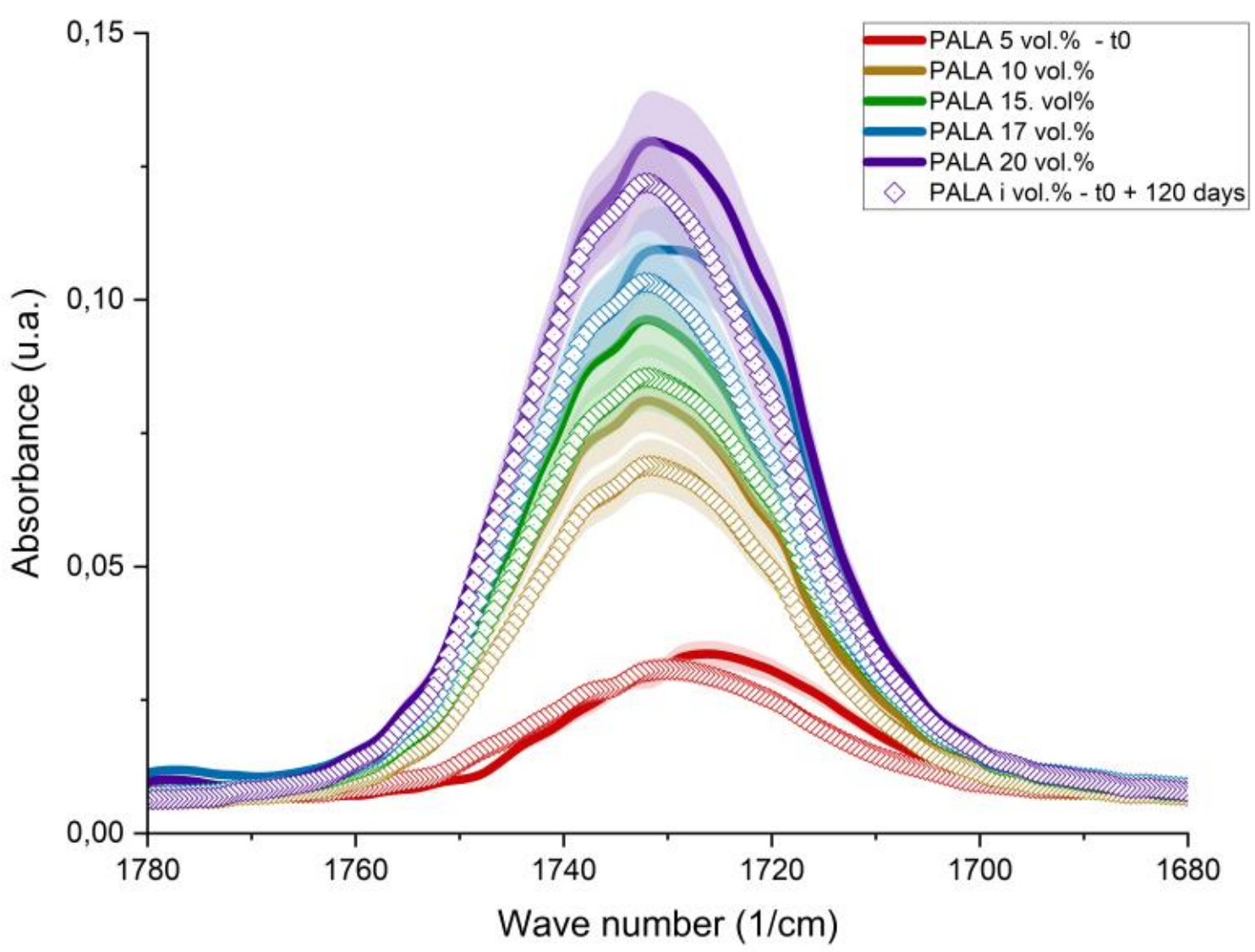

*Figure 2: Superposition of the IR spectra of plasticized films (5 to 20 vol.%) measured at $t_0$ (lines) and $t_0$+120 days (scatter) from 1780 to 1680 1/cm. The samples terpolymer x Palamoll i vol.% are hereafter referred to as xPALA i vol.%.*

Figure 2 shows the gradual reduction of the characteristic peak for Palamoll at 1730 1/cm over a period of four months in each film that was examined. The plain curves and the diamond-marked curves show the initial and second measurements respectively. Warm shades indicate the lowest Palamoll content, while colder shades represent the highest levels. The mean standard error of the absorbance over the wave number of a single peak was estimated to be of ± 7.3%. The decrease in peak intensity over time is visible but remains close or comprised within the order of magnitude of the margins of error of the measures. By measuring the area under each peak, it was estimated how much Palamoll migrated over time. According to Table 1, losses range from 4.3 ± 0.6 % to 14.7 ± 1.2 %, with fewest and highest losses observed for the 5 vol.% and 10 vol.% plasticized films respectively. It is not possible to reach a conclusion about the trend in percentage difference of plasticizer migration based on the Palamoll content. The overall observed losses remain below 16 %, confirming the strong chemical stability of the compound.

*Table 1: Evaluation of the actual Palamoll content in each of the plasticized films through a comparative analysis conducted by IR spectroscopy.*

| **Samples** | **Palamoll volume content (vol.%)**<br>**Initial** | **Palamoll volume content (vol.%)**<br>**120 days** | **Palamoll losses (%)** |
|---|---|---|---|
| **xPALA 5 vol.%** | 6.0 ± 0.4 | 5.7 ± 0.4 | 4.3 ± 0.6 |
| **xPALA 10 vol.%** | 12.8 ± 0.9 | 10.9 ± 0.8 | 14.7 ± 1.2 |
| **xPALA 15 vol.%** | 15.0 ± 1.1 | 13.6 ± 1.0 | 9.3 ± 1.5 |
| **xPALA 17 vol.%** | 17.5 ± 1.3 | 15.7 ± 1.1 | 10.2 ± 1.7 |
| **xPALA 20 vol.%** | 19.9 ± 1.5 | 18.1 ± 1.3 | 8.8 ± 2.0 |

## 3.2 Young Modulus in the kHz range

To confirm that the TTS principle applies to the neat and plasticized P(VDF-TrFE-CFE) samples, the constants C1 and C2 of the WLF model (Equation (11)) were determined and are gathered in Table 2, along with the constants of other common polymers found in literature. From Table 2 it appears that the WLF constants vary significantly from one polymer to another. This observation was acknowledged by the authors of the WLF model, stating that the uncertainties in specifying $T_g$ and the difficulty of experimental measurements near $T_g$ require careful consideration when applying Equation (11) [17]. Additionally, the universal coefficients C1=17.44 and C2=51.6 K derive from an averaging of data from different polymers, making Equation (11) not necessarily suitable for a particular system [20]. Here, the coefficients C1 and C2 obtained for neat and modified PVDF-TrFE-CFE samples, with C1 ranging from 2.8 ± 0.6 to 12 ± 2.4 and C2 from 55.2 ± 3.8 to 100.4 ± 13.8 K, fall within the ranges reported in the literature for polymers when the reference temperature differs from $T_g$ , confirming that these values are physically reasonable and enabling the use of the WLF model in this study [21], [22]. No specific trend is observed between the Palamoll content and the values of the constants.

*Table 2: Reported WLF constants C1 and C2 of various polymers with $T_g < T_0 < T_g + 100$. This work is compared to ref [21] and [22] .*

| | C1 | C2 (K) | Source |
|---|---|---|---|
| **Universal WLF constants** | 17.44 | 51.6 | [22] |
| **Polystyrene** | 12.7 | 48.9 | [21] |
| **Polyurethane** | 8.86 | 101.6 | [21] |
| **Polyisobutylene** | 8.61 | 200.4 | [21] |
| **Neat terpolymer** | 8.1 ± 0.8 | 69.6 ± 3.0 | This work |
| **xPALA 5 vol.%** | 8.4 ± 1.4 | 85.5 ± 6.6 | This work |
| **xPALA 10 vol.%** | 2.8 ± 0.6 | 55.2 ± 3.8 | This work |
| **xPALA 15 vol.%** | 12 ± 2.4 | 100.4 ± 13.8 | This work |
| **xPALA 17 vol.%** | 8.8 ± 1.4 | 89.3 ± 8.8 | This work |
| **xPALA 20 vol.%** | 9.0 ± 1.6 | 92.1 ± 10.6 | This work |

Elastic Moduli of pure and modified terpolymers with 5 to 20 vol.% of Palamoll in the kHz range is given on Figure 3(a). Lighter shades are associated with higher Palamoll content while the darker ones correspond to the lowest, with the black curve attributed to the pure terpolymer film. The progressive incorporation of Palamoll leads to a substantial and gradual decrease in the modulus of the pure terpolymer. On Figure 3(b) is displayed the Y' of the samples at 1 kHz in regard to their Palamoll volume content. With a 20 vol.% addition, the modulus is reduced by more than 80 %, declining from 335 MPa to 58.2 MPa. This observation is clearly associated with reduced crystallinity, which lowers the rigidity of the films when Palamoll is incorporated. Although increased flexibility is achieved, this is accompanied by a notable rise in $\tan(\delta)_m$ ,as shown on Figure 3(a) and (b). At 1 kHz, mechanical losses escalate from 25% to 61% upon the addition of 17–20 vol.% Palamoll. Increased mechanical losses lead to significant heat dissipation during actuation, thereby reducing electromechanical transduction efficiency [23]; this factor should be carefully evaluated in the assessment of overall system performance.

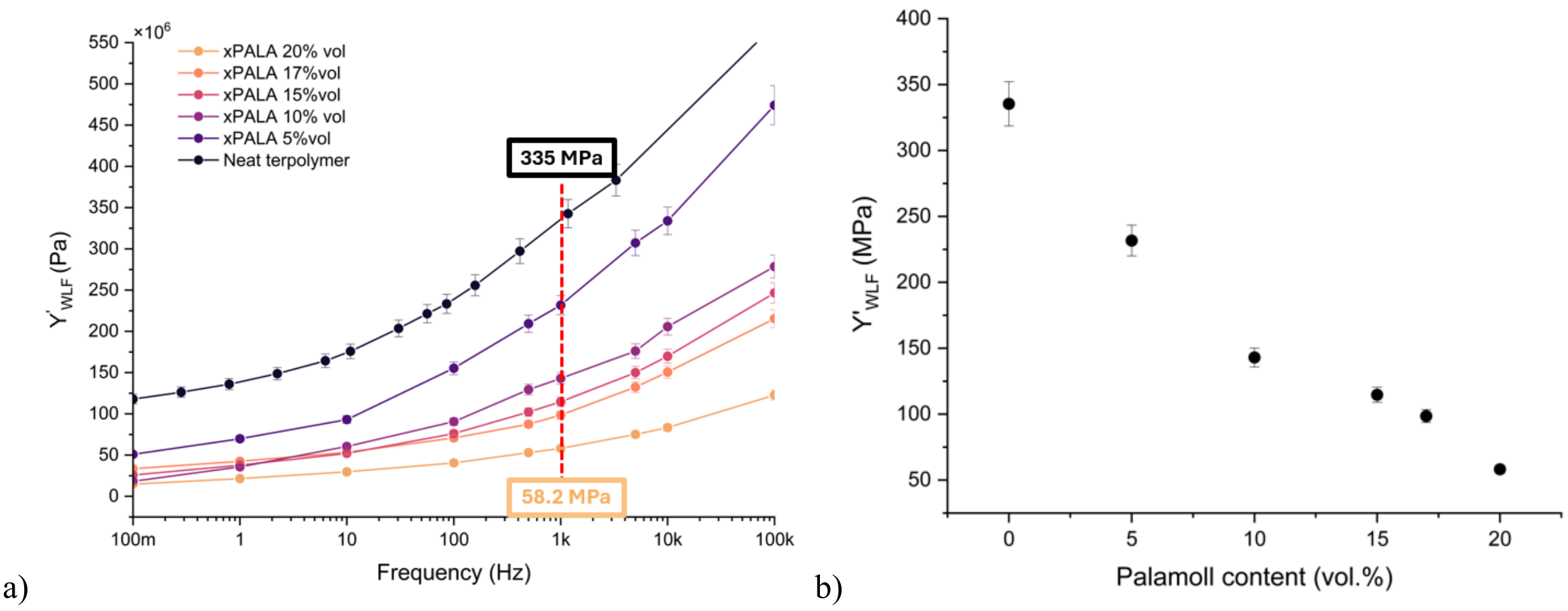

*Figure 3: Elastic Moduli (Y') extrapolated with WLF model over frequency (a) and as function of Palamoll content @1kHz (b) for neat and plasticized (5 to 20 vol.%) terpolymers.*

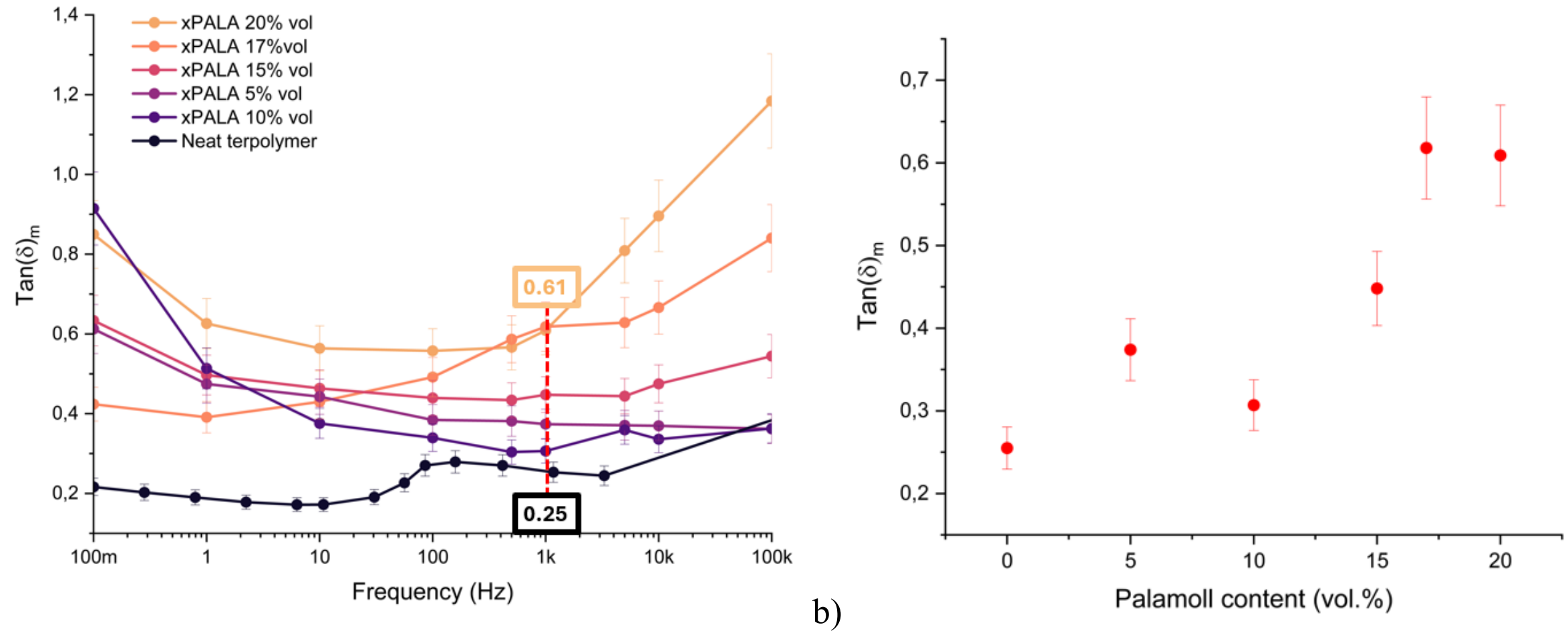

*Figure 4: Mechanical losses (tan(δ)ₘ) extrapolated with WLF model over frequency (a) and as function of Palamoll content @1kHz (b) for neat and plasticized (5 to 20 vol.%) terpolymers.*

## 3.3 Dielectric properties in the kHz range

At 1 kHz, dipolar orientation becomes the predominant polarization mechanism within the terpolymer, while interfacial polarization effects such as the Maxwell-Wagner-Sillars effect and ionic losses, typically observed between 0.01 and 0.1 Hz [4], are no longer significant. Subsequently, the current density of the conduction losses $J_l$ becomes neglectable as emphasized by Figure 5. The lighter shades on Figure 5(a) are associated with the films exhibiting the highest plasticizer content and the darker ones to the lowest contents, with the black curve attributed to the neat terpolymer film. For each of the pure and plasticized films, the total current density, plotted in red dots on Figure 5(a), matches the capacitive current density curves under 50 V/µm AC, emphasizing the pure capacitive behavior of the films as nearly no losses are observed (cf. Equation (6)).

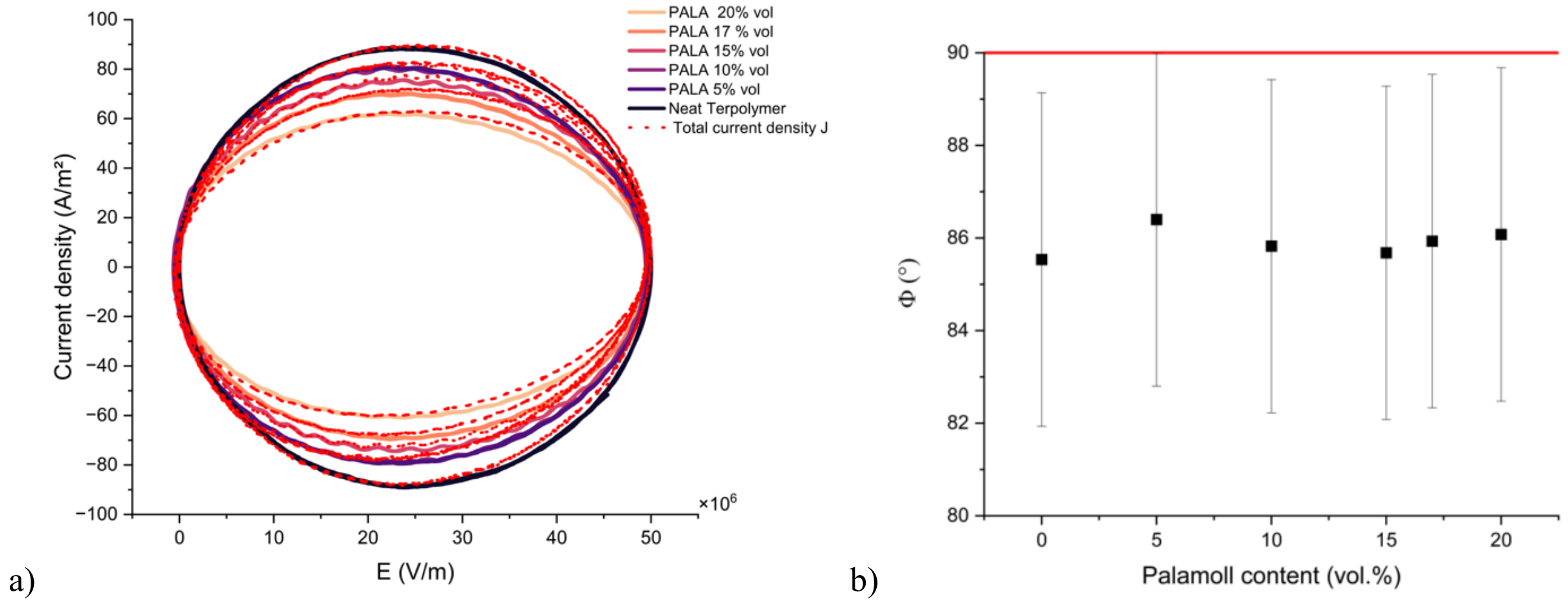

*Figure 5: (a) Total (red dots) and capacitive (lines) current densities of the neat and modified terpolymer under AC 50 V/µm field at 1 kHz; (b) Phase shift angle of the current response of the samples relative to the applied voltage at 1 kHz across 50 V/µm. The maximal physical reachable value is 90°, highlighted by the red line.*

Dielectric losses $tan(\delta)$ at 50 V/µm and 1 kHz have been further analyzed from the phase shift between the applied voltage and the current response of the films. The resulting phase angles of all the films have been found to range between 82 to 90°, error bars included, as depicted on Figure 5(b). Consequently, the capacitive behavior of the films can be confirmed as the very low shift indicates that the resistive in-phase current component is close to 0. To quantify the scope of the shift, the associated dielectric losses $\tan(\delta)$ are given on Figure 6(b). Figure 6(b) gathers both the results from BDS at low AC electric fields of ≈0.01 V/µm across $10^{-1}$ to $10^{5}$ Hz and from $\tan(\delta)$ at 50 V/µm and 1 kHz computed through Equation (9). Dielectric losses of the pure and plasticized films reach their minimum values near 1 kHz ranging from 6.4% to 8.2%, at both 0.01 and 50 V/µm. The concordance of the results confirms both the reliability of the two measurement approaches and indicates that the low-loss behavior is intrinsic to the terpolymer, remaining essentially unchanged across a wide range of electric fields. Values of $\tan(\delta)$ at 1 kHz and low AC electric field for each studied film are reported in Table 3.

From Figure 5(a) it is also clear that the capacitive current $J_C$ decreases gradually with the increasing addition of Palamoll up to 20 vol.%, piking at 88.9 and 61.4 A/m² for the neat and plasticized at 20 vol.% film respectively, which is explained by the decrease in real permittivity ε' through Equation (7). Similarly, at 50 V/µm and 1 kHz, the electric displacement D of the films decreases with the increasing Palamoll content as displayed on Figure 7(a).

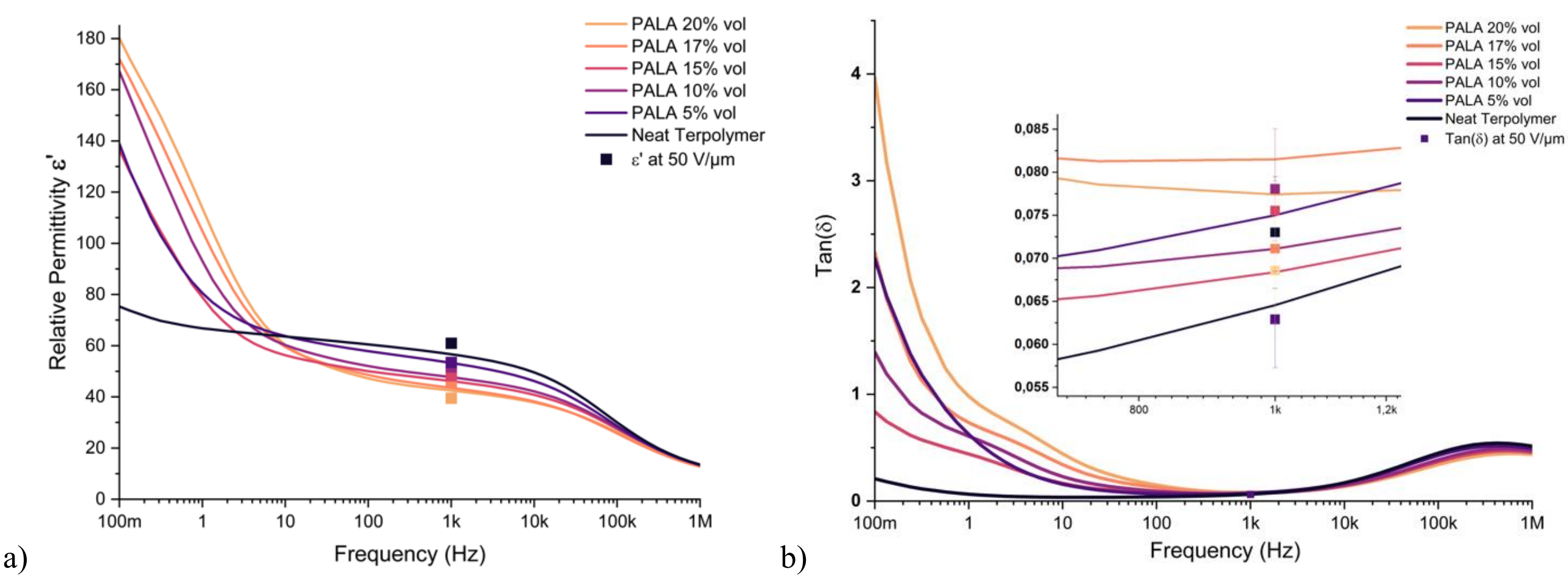

*Figure 6: (a) Relative permittivity of neat and modified terpolymer across frequency at 0.01 V/µm (curves) and at 50 V/µm (scatter); (b) dielectric losses over frequency of the neat and modified terpolymers at 0.01 V/µm with a focus at 1 kHz.*

*Table 3: Dielectric characteristics of the neat and modified terpolymers at 1 kHz. The values of the permittivities at 50 V/µm have been determined through the Debye-Langevin model and are the sum of $\varepsilon'_a$ $(= 12)$ and $\varepsilon'_c$ .*

| | **Relative Permittivity at low E ≈ 0.01 V/µm** | **Relative Permittivity at 50 V/µm $\varepsilon' = \varepsilon'_a + \varepsilon'_c$** | **Tan(δ) (%)** |
|---|---|---|---|
| **xPALA 5% vol** | 53.3 | 53.5 | 7.5 |
| **xPALA 10% vol** | 47.7 | 51.8 | 7.1 |
| **xPALA 15% vol** | 46.2 | 48.3 | 6.8 |
| **xPALA 17% vol** | 43.5 | 45.6 | 8.2 |
| **xPALA 20% vol** | 42.5 | 39.5 | 7.7 |
| **Neat Terpolymer** | 56.6 | 61.0 | 6.5 |

Figure 6(a) compares the real permittivity at 1 kHz obtained under a high AC field of 50 V/µm, calculated using the Debye–Langevin model, with the low-field BDS spectra (≈0.01 V/µm) recorded between $10^{-1}$ and $10^{5}$ Hz. The corresponding ε′ values at 1 kHz extracted from both approaches are summarized in Table 3. The close agreement between the low-field BDS measurements and the high-field permittivity predicted by the Debye–Langevin model supports the reliability of the latter for describing the dielectric response under strong electric fields. The decrease in ε′ at 1 kHz—from 61.0 to 39.5 at 50 V/µm and from 56.6 to 42.5 at 0.01 V/µm, from pure to 20 vol.% plasticized films —can be attributed to the increasing concentration of the non-polar Palamoll molecules, whose intrinsically low permittivity reduces the overall real permittivity of the blend in a proportional manner. Furthermore, incorporating the polymeric plasticizer results in a reduction of the terpolymer's crystallinity rate. This suggests that fewer crystalline dipoles undergo reorientation when subjected to an alternating electric field, which is directly evidenced by a decrease in ε′.

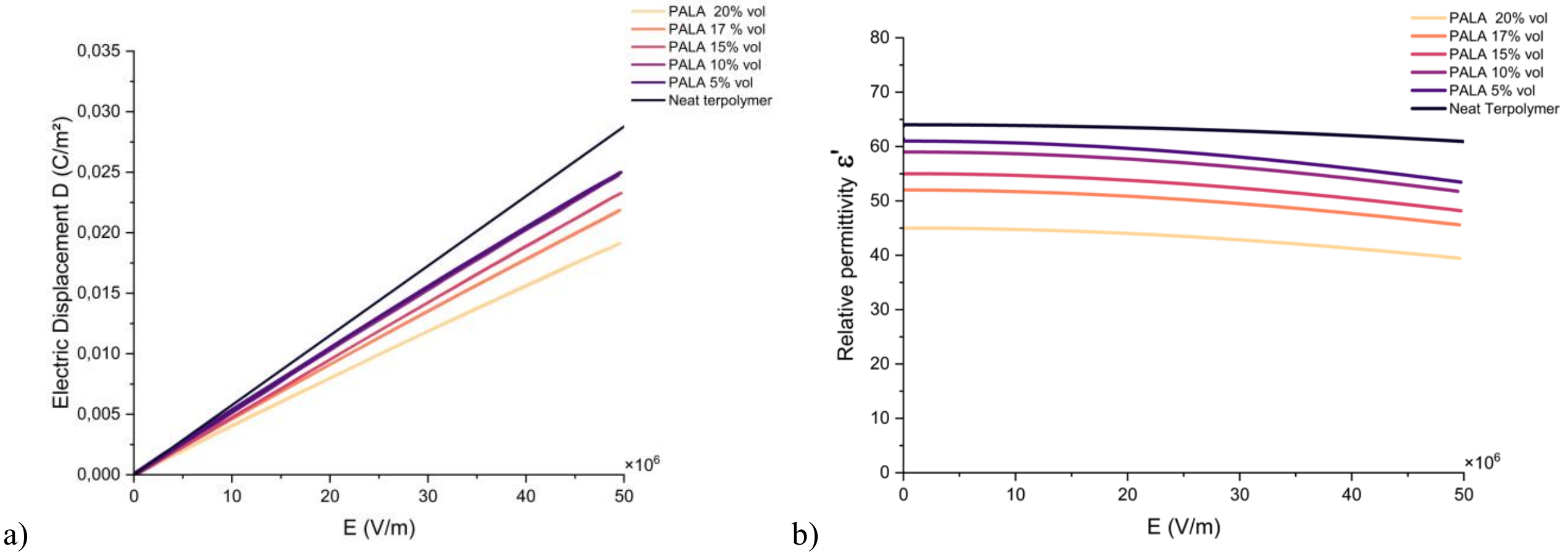

*Figure 7: (a) Electric Displacement obtained through the experimental $J_c$ and (b) Real Relative permittivity determined through the Debye-Langevin model of the pure and plasticized films at 1 kHz under 50 V/μm.*

Figure 7(b) shows the real permittivity at 1 kHz for both the neat and plasticized films as the electric field increases up to 50 V/μm. A slight decrease in ε′ is observed when the field rises above 10 V/μm, an effect that becomes more pronounced in the plasticized samples. This reduction is attributed to the electric field approaching the saturation field $E_S^c$ of the crystalline phases of each of the samples, particularly in the plasticized films where the saturation is reached more readily due to the enhanced dipole mobility. Through the fit of the experimental curves of $J_c(E)$ with $J_{c_{theo}}(E)$, $E_S^c$ was estimated to be of 60 V/μm for all the plasticized samples. Experimental fields of maximum 50 V/μm did not allow to find $E_S^c$ for the pure terpolymer film, consequently it was estimated to be equivalent to 100 V/μm in accordance with literature [10].

Dielectric breakdown strength results are displayed on Figure 8 and Weibull's parameters combined with $R^2$, the determination coefficient of the fit for each studied film, are gathered in Table 4. Although some variability is present, the $R^2$ values (0.94–0.98) indicate that the Weibull CFD model provides a consistent fit to experimental data. Looking at the characteristic breakdown field λ, it can be concluded that the addition of the polymeric plasticizer significantly reduces the electric field at which 63.2% of the samples break, dropping from 293 to 119 V/μm at the least, for neat and plasticized at 17 vol.% terpolymers respectively. However, higher plasticizer content leads to a narrower distribution, with the shape factor k equaling 14.6 at the highest, for a 20 vol.% plasticized film, against 3.97 for a neat terpolymer. The difference could be attributed to the heterogeneous structure of the semi-crystalline terpolymer compared to Palamoll. Since Palamoll is the limiting factor with a lower $E_b$, increasing the Palamoll rate could raise the breakdown concentration near $E_b$ of Palamoll.

For actuation applications, it is generally safer to use the lowest dielectric breakdown strength as the maximum allowable electric field. As the strain response of the material increases with $E^2$ (Equation (1)), lowering the operating electric field will significantly reduce the actuation strain output. Subsequently, a compromise will have to be found between the achievable strain output and a realistically safe operating electric field. However, drawing definitive conclusions regarding the lowest $E_b$ for each film requires a substantially larger number of tests, which falls outside the scope of this study.

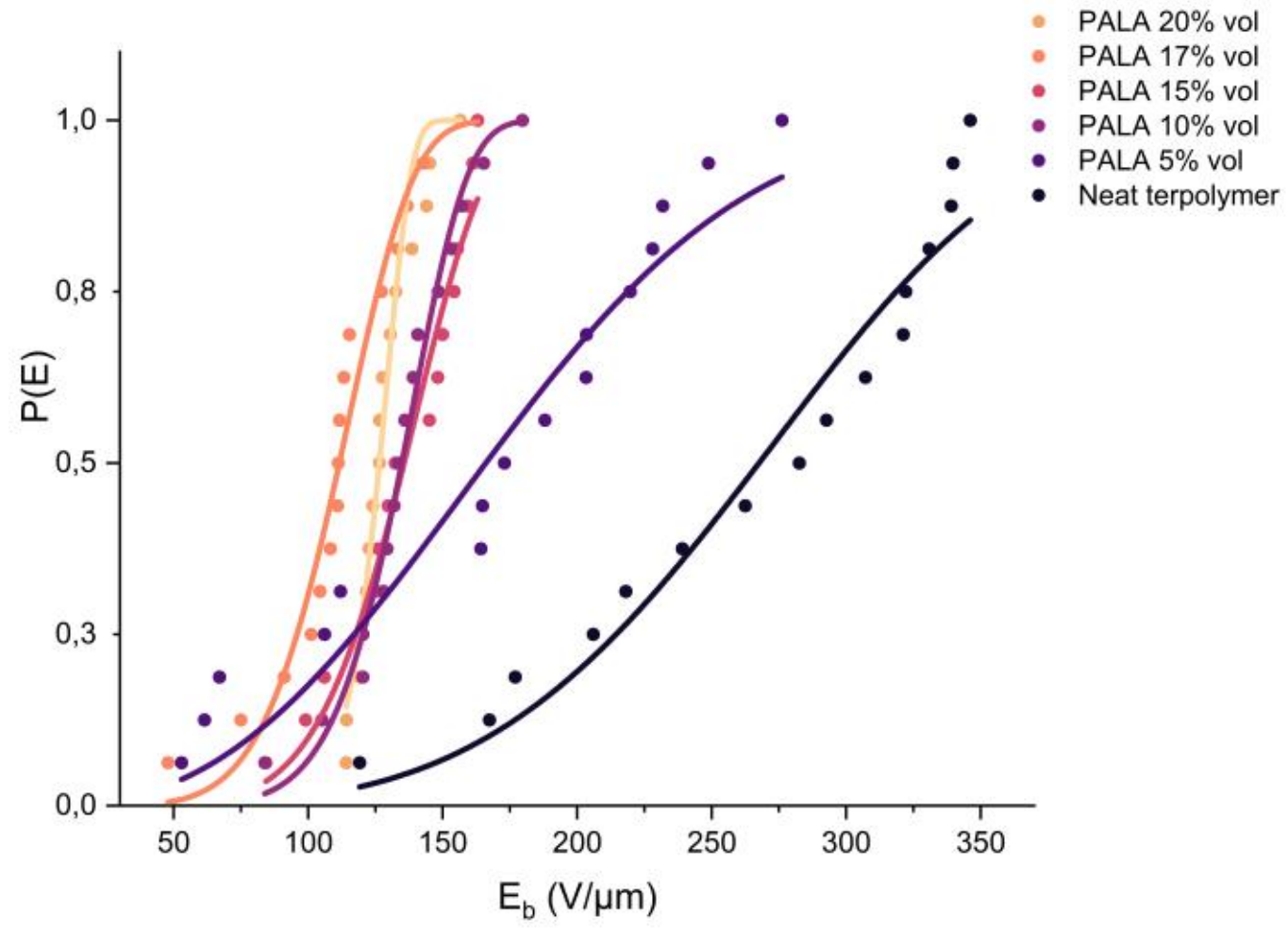

*Figure 8: Weibull distribution of the dielectric breakdown probability of neat and modified terpolymers.*

*Table 4: Parameters of the Weibull fit k and λ combined with the determination coefficient R² assessing the goodness of the fit for each of the studied films.*

| | **k** | **λ (V/µm)** | **R²** |
|---|---|---|---|
| **Neat terpolymer** | 3.97 | 293 | 0.95 |
| **xPALA 5 vol.%** | 2.52 | 192 | 0.94 |
| **xPALA 10 vol.%** | 7.73 | 141 | 0.98 |
| **xPALA 15 vol.%** | 6.20 | 144 | 0.97 |
| **xPALA 17 vol.%** | 5.78 | 119 | 0.96 |
| **xPALA 20 vol.%** | 14.6 | 130 | 0.96 |

### 3.4 Strain actuation performance

Strain actuation performance of the neat and plasticized samples was evaluated through the associated figure of merit; $FoM_{strain}$. The results are displayed on Figure 9(a) where $FoM_{strain}$ is plotted in regard to the content of Palamoll within the film, from 0 (neat terpolymer) to 20 vol.% at 1 kHz under 50 V/µm. Since $FoM_{strain}$ solely depends on ε' and Y' as detailed in Equation (3) with ε'(E) remaining nearly constant from 0.01 to 50 V/µm (cf. Figure 7(b)), the specific choice of the electric field does not significantly affect the outcome. Nonetheless, since the values of ε' reached at 50 V/µm are supposedly the lowest due to the proximity with the saturation field, the given FoMs represent the lower bound of the actual performances. Because the addition of Palamoll led to a major decrease in Y' up to 80 % and a maximum decrease of 30% of ε', the displayed results of Figure 9(a) rely mainly on the enhanced flexibility of the films. Subsequently, at a given electric field, $FoM_{strain}$ increases with the volume content of Palamoll. The highest $FoM_{strain}$ is attributed to the 20 vol.% plasticized film, reaching 6.00x10$^{-18}$ Fm/N, which is nearly four times greater than for the pure terpolymer, yielding 1.69x10$^{-18}$ Fm/N.

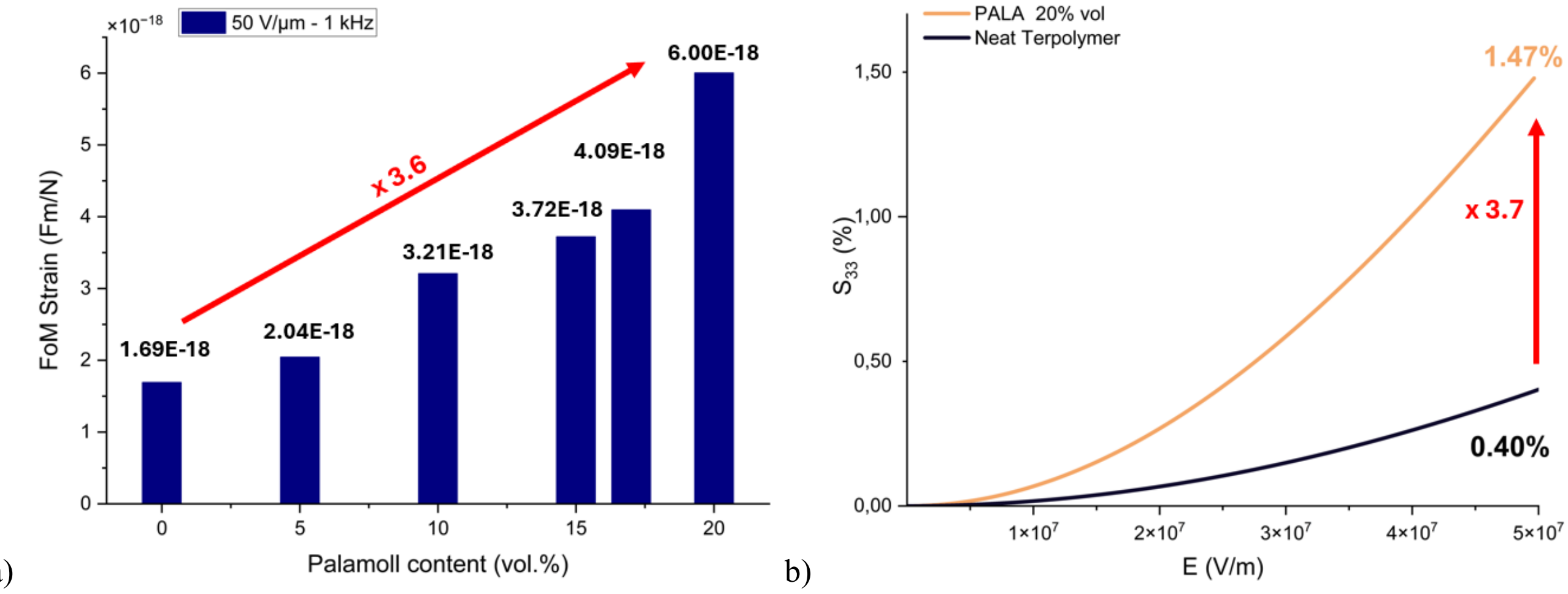

*Figure 9: (a) FoM strain of the neat and plasticized (5 to 20 vol.%) films computed with ε' from the Debye-Langevin model at 50 V/µm and Y' from the WLF model, both taken at 1 kHz; (b) Estimated strain output (%) of the neat and plasticized at 20 vol.% films under a maximum operating electric field of 50 V/µm at 1 kHz.*

On Figure 9(b) are displayed the strain outputs $S_{33}$ of the neat and 20 vol.% plasticized films evaluated at a maximum AC operating electric field of 50 V/µm at 1 kHz in accordance with FlexSiMirror requirements. The gradual addition of Palamoll leads to a rise of $S_{33}$ with a major 3.7-fold increase relative to the neat terpolymer, observed for the film containing 20 vol.% plasticizer at 50 V/µm. Specifically, films of 100 µm thick undergoing 50 V/µm at 1 kHz can provide 1.5 µm transversal deformation when plasticized at 20 vol.% against 0.4 µm for the pure terpolymer.
When it comes to designing actuators in the context of FlexSiMirror project, it can be concluded that 3.7 times less layers will be required with modified films of PVDF-TrFE-CFE with 20 vol.% Palamoll compared with the pure material. However, as the structure of an actuator is embedded on the top and bottom surfaces and consists of a succession of the electroactive polymer (EAP) layers with an electrode (cf. Figure 1), the total amount layers required for the application cannot be determined through the sole strain output value of a free EAP layer. Additionally, the deformation of the actuator pillar cannot be inferred solely from the nominal field strength (e.g., 50 V/µm). The maximum supply voltage that can be applied to the stacked structure must be defined beforehand, otherwise no reliable conclusion can be drawn regarding the overall displacement.

## 4. CONCLUSION

The dielectric and mechanical characterization of neat PVDF-TrFE-CFE and modified films with 5 to 20 vol.% of Palamoll revealed that actuation strain in the kHz can be enhanced by up to a factor 3.6. This improvement arises from the incorporation of up to 20 vol.% Palamoll, which reduces the elastic modulus at 1 kHz by more than fivefold without impacting significantly the permittivity, thereby boosting the strain-related figure of merit by over threefold. Not only the modified films exhibit good chemical compatibility with the terpolymer matrix, with Palamoll migration limited to 8.8 ± 2.0 % at 20 vol.% over 4 months, but dielectric losses measured at 50 V/µm AC and 1 kHz were found to be minimal, reaching a maximum of 8.4%. These findings demonstrate that the modified terpolymer actuators can perform well and have a long service life since minimal leakage current significantly reduces the chances of energy loss and self-heating, thereby supporting strong dielectric reliability [24]. Further work regarding the dielectric characterization includes conducting polarization cycles at higher fields, nearer to the saturation field, as it may yield more precise data for applying the Debye-Langevin model to both the pure and plasticized terpolymer. Additionally, increasing the number of samples when performing dielectric strength tests could provide more accurate results of $E_b$ and improve the quality of the Weibull CFD fit. This is of critical importance as the experimental values found for $E_b$ dictates the expected strain outputs during actuation. Moreover, a study on the interaction between Palamoll molecules and the polymer matrices should be performed to better analyze the most suitable Palamoll content required for the application.

## ACKNOWLEDGEMENTS

The authors acknowledge support from the PEPR Origins (WP1.2 – FlexSiMirror – ANR-22-EXOR-0004) for its financial support within the Plan France 2030 of the French government operated by the National Research Agency (ANR). Partners of the project are the research astrophysics center of Lyon CRAL/CNRS, the microelectronic center of Provence (CMP) with Les Mines de St-Etienne and the astrophysics laboratory of Marseille LAM/CNRS.

## REFERENCES

[1] P.-H. Ducrot, I. Dufour, et C. Ayela, « Optimization Of PVDF-TrFE Processing Conditions For The Fabrication Of Organic MEMS Resonators », *Sci. Rep.*, vol. 6, nº 1, p. 19426, janv. 2016, doi: 10.1038/srep19426.

[2] N. Della Schiava *et al.*, « Enhanced Figures of Merit for a High-Performing Actuator in Electrostrictive Materials », *Polymers*, vol. 10, nº 3, p. 263, mars 2018, doi: 10.3390/polym10030263.

[3] B. Qiao, X. Wang, S. Tan, W. Zhu, et Z. Zhang, « Synergistic Effects of Maxwell Stress and Electrostriction in Electromechanical Properties of Poly(vinylidene fluoride)-Based Ferroelectric Polymers », *Macromolecules*, vol. 52, nº 22, p. 9000-9011, nov. 2019, doi: 10.1021/acs.macromol.9b01580.

[4] N. Della Schiava, M. Le, J. Galineau, F. Domingues Dos Santos, P. Cottinet, et J. Capsal, « Influence of Plasticizers on the Electromechanical Behavior of a P(VDF-TrFE-CTFE) Terpolymer: Toward a High Performance of Electrostrictive Blends », *J. Polym. Sci. Part B Polym. Phys.*, vol. 55, nº 4, p. 355-369, févr. 2017, doi: 10.1002/polb.24280.

[5] J.-F. Capsal, J. Galineau, M.-Q. Le, F. Domingues Dos Santos, et P.-J. Cottinet, « Enhanced electrostriction based on plasticized relaxor ferroelectric P(VDF-TrFE-CFE/CTFE) blends », *J. Polym. Sci. Part B Polym. Phys.*, vol. 53, nº 19, p. 1368-1379, oct. 2015, doi: 10.1002/polb.23776.

[6] Hunter, Ian W.; Madden, John D.; Vandesteeg, Nate; Madden, Peter G.; Takshi, Arash, « Artificial Muscle Technology: Physical Principles and Naval Prospects », MASSACHUSETTS INST OF TECH CAMBRIDGE OFFICE OF SPONSORED RESEARCH, 2003.

[7] Kritsadi Thetpraphi, *Development of electroactive polymer actuators for next generation mirror : Live-Mirror.* Université de Lyon, 2020.

[8] BUREAU Jean-Marc, « Propriétés diélectriques des Polymères », *Tech. Ing.*, p. 24, févr. 2016.

[9] B. Chu *et al.*, « A Dielectric Polymer with High Electric Energy Density and Fast Discharge Speed », *Science*, vol. 313, nº 5785, p. 334-336, juill. 2006, doi: 10.1126/science.1127798.

[10] J.-F. Capsal, M. Lallart, J. Galineau, P.-J. Cottinet, G. Sebald, et D. Guyomar, « Evaluation of macroscopic polarization and actuation abilities of electrostrictive dipolar polymers using the microscopic Debye/Langevin formalism », *J. Phys. Appl. Phys.*, vol. 45, nº 20, p. 205401, mai 2012, doi: 10.1088/0022-3727/45/20/205401.

[11] F. Pedroli, A. Marrani, M. Le, C. Froidefond, P. Cottinet, et J. Capsal, « Processing optimization: A way to improve the ionic conductivity and dielectric loss of electroactive polymers », *J. Polym. Sci. Part B Polym. Phys.*, vol. 56, nº 16, p. 1164-1173, août 2018, doi: 10.1002/polb.24636.

[12] S.-L. Wang, « On the definitions of practical permittivity and dielectric loss angle », *J. Frankl. Inst.*, vol. 326, nº 2, p. 247-254, janv. 1989, doi: 10.1016/0016-0032(89)90072-0.

[13] X. Zhou *et al.*, « Electrical breakdown and ultrahigh electrical energy density in poly(vinylidene fluoride-hexafluoropropylene) copolymer », *Appl. Phys. Lett.*, vol. 94, nº 16, p. 162901, avr. 2009, doi: 10.1063/1.3123001.

[14] M. A. Meyers et K. K. Chawla, *Mechanical behavior of materials*, 2. ed., 4. print. with corr. Cambridge: Cambridge University Press, 2010.

[15] C. Smithson, M. Stamenović, M. Nujkić, et S. Putić, « Time: Temperature superposition principle: Application of WLF equation in polymer analysis and composites », *Zastita Mater.*, vol. 55, nº 4, p. 395-400, déc. 2014, doi: 10.5937/ZasMat1404395L.

[16] H.-M. Bao, J.-F. Song, J. Zhang, Q.-D. Shen, C.-Z. Yang, et Q. M. Zhang, « Phase Transitions and Ferroelectric Relaxor Behavior in P(VDF−TrFE−CFE) Terpolymers », *Macromolecules*, vol. 40, nº 7, p. 2371-2379, avr. 2007, doi: 10.1021/ma062800l.

[17] M. L. Williams, R. F. Landel, et J. D. Ferry, « The Temperature Dependence of Relaxation Mechanisms in Amorphous Polymers and Other Glass-forming Liquids », *J. Am. Chem. Soc.*, vol. 77, nº 14, p. 3701-3707, juill. 1955, doi: 10.1021/ja01619a008.

[18] C. Lesenne, D. Audigier, et J. Capsal, « Physico-Chemical Understanding of Plasticizers Interaction with P(VDF-TrFE-CFE) Electroactive Polymer », *Macromol. Mater. Eng.*, p. e00275, oct. 2025, doi: 10.1002/mame.202500275.
[19] Yong Wang, Sheng-Guo Lu, M. Lanagan, et Qiming Zhang, « Dielectric relaxation of relaxor ferroelectric P(VDF-TrFE-CFE) terpolymer over broad frequency range », *IEEE Trans. Ultrason. Ferroelectr. Freq. Control*, vol. 56, nº 3, p. 444-449, mars 2009, doi: 10.1109/TUFFC.2009.1063.
[20] M. Peleg, « On the use of the WLF model in polymers and foods », *Crit. Rev. Food Sci. Nutr.*, vol. 32, nº 1, p. 59-66, janv. 1992, doi: 10.1080/10408399209527580.
[21] Ferry, J. D., « Viscoelastic properties of polymers ». John Wiley & Sons, 1980.
[22] R. P. White et J. E. G. Lipson, « Polymer Free Volume and Its Connection to the Glass Transition », *Macromolecules*, vol. 49, nº 11, p. 3987-4007, juin 2016, doi: 10.1021/acs.macromol.6b00215.
[23] G. Gallucci et A. Hunt, « Poly(vinylidene Fluoride)-Based Ferroelectric Polymers for Electromechanical Transduction: A Systematic Review of Materials and Actuators », *Adv. Intell. Syst.*, p. e202500694, déc. 2025, doi: 10.1002/aisy.202500694.
[24] F. Pedroli, A. Marrani, M.-Q. Le, O. Sanseau, P.-J. Cottinet, et J.-F. Capsal, « Reducing leakage current and dielectric losses of electroactive polymers through electro-annealing for high-voltage actuation », *RSC Adv.*, vol. 9, nº 23, p. 12823-12835, 2019, doi: 10.1039/C9RA01469A.